\documentclass[conference,a4paper]{IEEEtran}

\IEEEoverridecommandlockouts
\usepackage{cite}
\usepackage{amsmath,amssymb,amsfonts}
\usepackage{algorithm}
\usepackage[noend]{algpseudocode}
\usepackage{graphicx}
\usepackage[caption=false]{subfig}
\usepackage{textcomp}
\usepackage{xcolor}
\usepackage{diagbox}
\usepackage{fixme}
\fxsetup{
    status=draft,
    margin=true
}

\newcommand{\bmv}{\texttt{bmv2}}

\def\BibTeX{{\rm B\kern-.05em{\sc i\kern-.025em b}\kern-.08em
    T\kern-.1667em\lower.7ex\hbox{E}\kern-.125emX}}

\begin{document}

\title{A Case for Data Centre Traffic Management on Software Programmable Ethernet Switches}

\author{

	\IEEEauthorblockN{
	    Kamil Tokmakov
	}
	\IEEEauthorblockA{High Performance Computing Center Stuttgart (HLRS) \\
	Stuttgart, Germany\\
	Email: tokmakov@hlrs.de}

	\and

	\IEEEauthorblockN{
	    Mitalee Sarker,
	    J\"org Domaschka and
	    Stefan Wesner
	}
	\IEEEauthorblockA{Institute of Information Resource Management\\
	Ulm University, Germany\\
	Email: firstname.lastname@uni-ulm.de}
}

\newpage

\vfill

\twocolumn[
\centering
\vspace{25em}
\LARGE
Copyright \copyright~2019 IEEE \\
\copyright~2019 IEEE. Personal use of this material is permitted. Permission
from IEEE must be obtained for all other uses, in any current or future
media, including reprinting/republishing this material for advertising or
promotional purposes, creating new collective works, for resale or
redistribution to servers or lists, or reuse of any copyrighted
component of this work in other works.
]

\vfill

\clearpage

\IEEEoverridecommandlockouts
\IEEEpubid{\makebox[\columnwidth]{978-1-7281-4832-8/19/\$31.00~\copyright2019 IEEE \hfill} \hspace{\columnsep}\makebox[\columnwidth]{ }}

\maketitle

\IEEEpubidadjcol

\begin{abstract}
Virtualisation first and cloud computing later has led to a consolidation of workload in data centres that also comprises latency-sensitive application domains such as High Performance Computing and telecommunication. These types of applications require strict latency guarantees to maintain their Quality of Service. In virtualised environments with their churn, this demands for adaptability and flexibility to satisfy. At the same time, the mere scale of the infrastructures favours commodity (Ethernet) over specialised (Infiniband) hardware.
For that purpose, this paper introduces a novel traffic management algorithm that combines Rate-limited Strict Priority and Deficit round-robin for latency-aware and fair scheduling respectively. In addition, we present an implementation of this algorithm on the \bmv{} P4 software switch by evaluating it against standard priority-based and best-effort scheduling.
\end{abstract}


\section{Introduction}

Virtualisation first and cloud computing later have led to a consolidation of workload in data centres. This on the one hand, reduces the amount of required servers and saves cost. Yet, on the other hand, it also makes it more difficult to provide resource guarantees to individual applications and chains of applications. Existing approaches have largely focused on resource provisioning with respect to CPU and disk; aspects regarding the network have found less attention~\cite{popa2012faircloud}.

Hence, the challenge of how to satisfy their dedicated demands when moving network-sensitive applications into virtualised infrastructures is still present: A large class of applications from the domain of high performance computing~(HPC) requires very low and predictable network latency. Classically, HPC data centres solve this issue by providing dedicated, expensive fibre and specific protocols such as Infiniband and Omnipath. In contrast, cloud data centres mostly rely on much cheaper commodity Ethernet wiring. Similar issues occur in future 5G networks, where network slicing and the virtual Evolved Packet Core~(vEPC) require support for latency-sensitive applications.

While off-the-shelf Ethernet is not able to achieve low latency or prioritisation of traffic, there exist strategies for traffic management that can mitigate delay for high priority traffic. Yet, those approaches are either very inflexible to set-up~\cite{kim2013improving} or do not provide the required sophistication to entirely avoid starvation for low priority traffic~\cite{wadekar2007enhanced}. In particular, programming the data plane through the widely used and adopted OpenFlow standard is not powerful enough, as OpenFlow's finite set of hardware-independent match and action pipelines do not support the creation of customised traffic management algorithms.

Addressing these challenges, this paper makes the following contributions: \emph{(i)} we introduce a novel traffic management algorithm by combining Rate-limited Strict Priority (RL-SP) and Deficit round-robin (DRR). Our RL-SP-DRR ensures a minimum amount of fairness, but still supports demands for ultra-low latency in Ethernet-based scenarios such as High-performance Computing and traffic-aware edge computing. \emph{(ii)} We present an implementation and P4-based configuration of this algorithm in the \bmv{}\footnote{https://github.com/p4lang/switch} P4 software switch. Doing so, we show the capability to realise custom traffic management based on standardised interfaces. \emph{(iii)} We evaluate this algorithm on \bmv{} by comparing it against standard priority-based and best-effort scheduling.

The remainder of this document is structured as follows. Section~\ref{sec:bg} introduces basic traffic management concepts and discusses related work. Section~\ref{sec:approach} is concerned with our approach, its design and realisation. Finally, Section~\ref{sec:eval} presents our evaluation and discusses its results before we conclude in the final section.

\section{Background}
\label{sec:bg}

\subsection{Software-defined Networking}
Software Defined Networking (SDN) postulates a programmability concept for network devices that decouples control plane and data plane. In consequence, forwarding decisions are not limited to existing routing algorithms, but can be set by the control plane. Thus, the complexity of the forwarding devices decreases, providing only the environment and interfaces for its programmability. Standardising such programmability avoids negative effect of vendor lock-in, leading to vendor independence in the heterogeneous infrastructure as cloud data centres have~\cite{azodolmolky2013cloud}.

Being a concept, SDN is realised through multiple protocols, open standards, and implementation methods~\cite{kreutz2015software}, with OpenFlow~\cite{mckeown2008openflow} as the most widely used open, standardised protocol. P4~\cite{bosshart2014p4} is a programming language and runtime-specification which promises through its programming approach to overcome OpenFlow's finite set of hardware-independent match and action pipelines. P4 offers user-defined packet processing and header parsing which leads to arbitrary match and action pipelines.

\subsection{Traffic Management}

In order to deal with the traffic excesses and practically implement
quality-of-service (QoS) guarantees (e.g. throughput, transfer delay, packet
losses), traffic management (TM) offers an access
control~\cite{chen2007network}, which assures that the traffic admitted to the
network conforms defined QoS guarantees by regulating the rate of the traffic
flow. Traffic policing and traffic shaping refer to the methods for such rate
regulation~\cite{blake1998architecture}. Traffic policing either drops excess
of packets or marks them with congestion experience flag once the rate of a
traffic flow reaches some threshold. In contrast, traffic shaping exploits
queueing mechanism of the forwarding devices in order to buffer traffic bursts
and then processes them at the desired rate smoothing the traffic. A queueing
mechanism is also useful for packet scheduling, e.g., when multiple packets of
different queues compete for the same output port of a forwarding device.
Here, the scheduler decides which queue to serve first. The packet scheduling
algorithms applied for those tasks allow for various prioritisation and fair
schemes.

Rate-limited Strict Priority (RL-SP)~\cite{li1997static} algorithm admits a
priority traffic with a specified rate and the remaining rates will be
distributed to the lower priorities without causing their complete starvation.
However, in the absence of prioritised traffic, low priority traffic will
still undergo the same rate distribution underutilising the link capacity. To
resolve this problem, weighted fair algorithms were proposed, which have rate
distributions based on relative weights. Weighted fair queueing
(WFQ)~\cite{demers1989analysis} and weighted round-robin
(WRR)~\cite{katevenis1991weighted} can provide higher bandwidth share to the
priority traffic while utilising the link in its absence, however, WFQ suffers
from computation complexity and WRR deviates rate distribution for variable
length of the packets. Deficit round-robin (DRR)~\cite{shreedhar1996efficient}
resolves these issues by introducing quantum values (weights) and deficit
counter (number of remaining bytes to be passed) for each queue, but its
drawback is round delays, since the queues are still served in round-robin
fashion. These delays cause latency for prioritised traffic, even if the
higher bandwidth is allocated. Table \ref{tbl:sched-sum} summarises most
common packet scheduling algorithms.

\begin{table*}
 	\caption{Summary of the packet scheduling algorithms}
	\label{tbl:sched-sum}
	\centering
	\begin{tabular}{|m{15em}|m{15em}|m{14em}|m{14em}|}
		
		\hline
		Algorithm & Principle & Advantage & Disadvantage \\
		\hline
		Rate-limited Strict Priority (RL-SP) & 
		Serve higher priority until some rate & 
		No starvation for lower priority & 
		Underutilisation  \\
		\hline
		Weighted fair queueing (WFQ) & 
		Serve a queue with lowest virtual finish time & 
		Fairness & 
		Calculation complexity \\
		\hline
		Weighted round-robin (WRR) & 
		Round robin based on weights & 
		Low complexity \& fair for fixed length packets & 
		Round delays \& unfair bandwidth share for variable length packets\\
		\hline
		Deficit round-robin (DRR) & 
		Serve a queue until deficit counter is sufficient & 
		Low complexity \& fair for variable length packets & 
		Round delays \\
		\hline
		
	\end{tabular}
\end{table*}

\subsection{Related Work}
\label{sec:bg-related}

\cite{hauser2017dynamic} proposes a centralised dynamic scheduler with
OpenFlow, responsible for fair sharing of network resources. The bandwidth 
estimation is done on the OpenFlow controller, hence the proximity to the
controller affects the performance of the scheduler, resulting in higher
reaction time and lower accuracy in utilisation. Moreover, the scheduling
safeguards high priority traffic from malicious traffic, but does not
explicitly prioritise the traffic in terms of higher bandwidth and lower
latency. The DRR in our approach results in more accurate utilisation due to
implicit rate distribution, and provides prioritisation needed for
latency-critical services.

QoSFlow~\cite{ishimori2013control} detects the inflexibility of OpenFlow for the selection of packet scheduling techniques and wrapped around a framework to resolve this. QoSFlow uses \texttt{netlink} sockets for interprocess communication between Linux user and kernel space, and queueing disciplines of the kernel to provide control over the packet scheduling algorithms. They extended OpenFlow 1.0 and datapath with queue
discipline selection messages to allow the kernel to appropriately schedule the
packets and enable configurable QoS in the datapath. Such customisation
requires the extension of OpenFlow and data plane, straying
from the standardisation process of OpenFlow and its vendor adoption. In contrast the P4-based
customisation used in this paper are easier to achieve due to the user-defined nature of data plane and auto-generated control plane API.

\cite{rad2015low} utilises Infiniband interconnects in the OpenStack
cloud platform to provide low-latency communication in the cloud
infrastructure.  
The results show that the performance is comparable to the bare-metal deployment. However, features such as live migration and checkpointing are challenging to provide for Infiniband virtualisation~\cite{mauch2013high}. Another downside of that approach is the greater cost of equipment.

\section{Design and Implementation}
\label{sec:approach}

This section first summarises the requirements imposed on a flexible traffic management system. Based on these requirements, we present our conceptual approach that combines strict priority scheduling with rate limitation (RL-SP) as a prioritisation mechanism and deficit round-robin (DRR). After that, we discuss how this algorithm is realised technically on the P4 switch and configured using the P4 language.

\subsection{Requirements}

The desire to apply dynamic and priority-based network scheduling in highly volatile and dynamically changing environments leads to a set of requirements towards the execution environment and the technical capabilities of the platform. We summarise three of these and also discuss the impact each particular requirement has on the realisation.

\noindent{\textbf{R.1~(Adaptability):}} Multi-tenancy environments are highly dynamic in nature and the traffic management needs to be able to cope with churned workloads, e.g. virtual machines coming and going. Furthermore, the workload mix may change over time so that the traffic management may require to be changed at runtime.

\noindent{\textbf{R.2~(Fairness)}} Fairness guarantees that no traffic in the network is experiencing starvation. In consequence, this means that all flows passing through a switch shall eventually make progress. Obviously, the progress of low priority traffic should be significantly less than the progress of high priority traffic.

\noindent{\textbf{R.3~(Transparency)}} Considering the setting of a virtualised infrastructure, it is important that the traffic management does not require changes on the level of operating systems and virtual machines. Hence, for the sake of user comfort, traffic management should be able to work with standard protocols and support the default user stack.

\subsection{Conceptual Approach}

\begin{figure}
	\centering
	\includegraphics[width=\columnwidth]{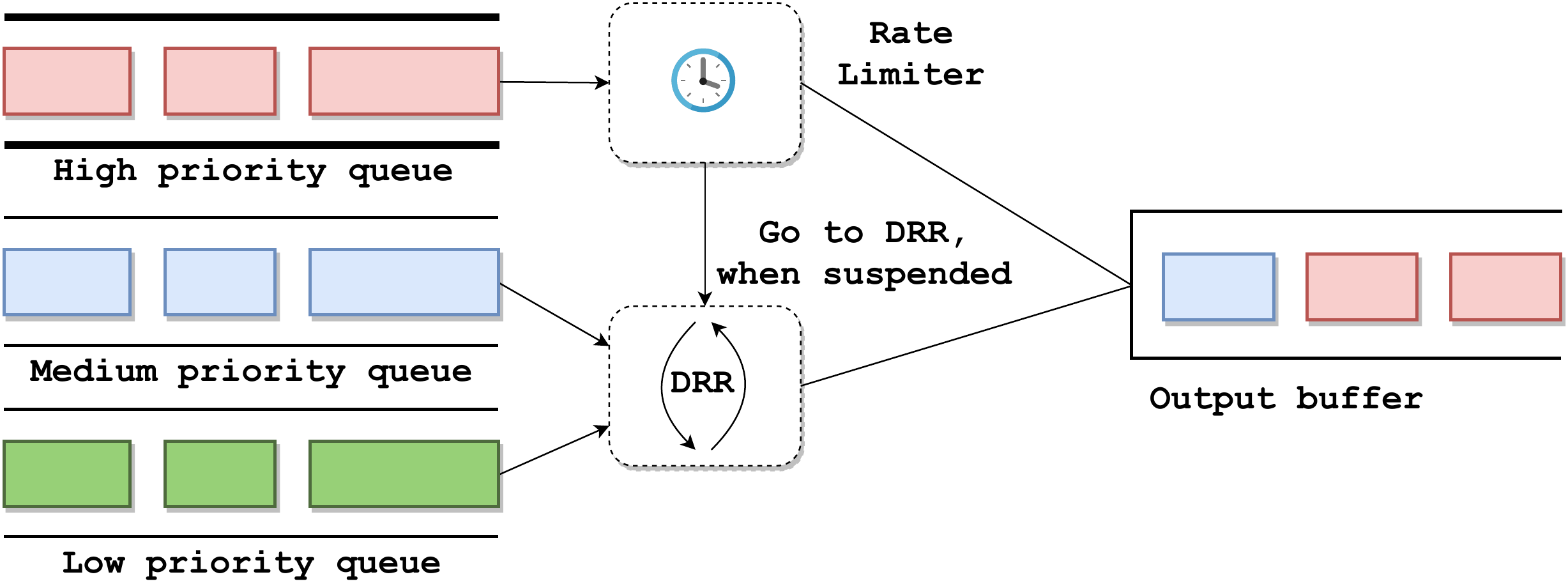}
	\caption{Rate limited strict priority with DRR (RL-SP-DRR)}
	\label{fig:impl-rl-sp-drr}
\end{figure}

The current state of SDN-enabled devices allow basic configuration of traffic
management parameters via the programmable control plane. Yet, with respect
to packet scheduling, only a queue selection for a particular flow is
supported. The scheduling itself is performed by the algorithms pre-installed
by a hardware manufacturer and configured via the interfaces supplied by the
manufacturer. This imposes a vendor lock-in issue in terms of
traffic management programmability. 

To overcome this issue, the control plane must provide more sophisticated and
customisable application programmable interfaces (APIs) to configure packet
scheduling parameters. While OpenFlow satisfies \textit{Adaptability~(R.1)}, its 
finite set of match and actions is a drawback, as any provision of traffic management features will require changes to the protocol itself, thus breaking the \textit{Transparency~(R.3)} requirement.

P4-based approaches on the other hand, auto-generate the control plane API during the
compilation process of a P4 program and adjust it according to the workflow and
pipeline definitions of the program. This enables the necessary configurations of the
packet scheduling parameters via control plane, while following standard user stack and protocols. For this reason, we use a P4-based approach implementing a custom traffic management algorithm~(RL-SP-DRR) in order to address aforementioned requirements.

RL-SP-DRR combines strict priority scheduling with rate limitation (RL-SP) as a prioritisation mechanism and deficit round-robin (DRR) as a fair scheduler. As depicted in Figure~\ref{fig:impl-rl-sp-drr} a high priority queue (HPQ) is reserved for strict prioritisation, so that it will be always served first, and lower priority queues (LPQs) will be served next. The HPQ occupies a dedicated rate of a link and the remaining link capacity will be distributed among LPQ based on their DRR's quantum values. This approach ensures \textit{Fairness~(R.2)}. In the absence of high priority traffic, DRR takes over the packet scheduling utilising the link capacity. Therefore, no involvement of the control plane is needed to re-establish the bandwidth allocation.

\subsection{Technical Approach}

\begin{algorithm}
	\begin{algorithmic}[1]
		\small
		\State $N$ \Comment Number of queues
		\State $queues[0..N-1]$ \Comment Priority queues
		\State $active\_list[0..N-2]$ \Comment List of active low priority queues
		\State $HPQ = queues[0]$ \Comment High priority queue

		\Procedure{Enqueue}{$q$, $p$}
		\If {$q$ $is$ $HPQ$}
		\State $p.departure = q.last\_sent + p.size/q.rate$
		\Else
		\State $p.departure = current\_time$
		\If {$q$ $is$ $empty$}
		\State $active\_list.push(q)$
		\EndIf
		\EndIf
		\EndProcedure

		\Procedure{Dequeue}{}
		\If {$HPQ.head.departure \le current\_time$}
		\State $Send\_Packet\_From(HPQ)$
		\Else
		\State $Perform\_DRR(active\_list)$
		\EndIf
		\EndProcedure

	\end{algorithmic}
	\caption{Rate-Limited Strict Priority with Deficit Round-Robin (RL-SP-DRR)}
	\label{algo:rl-sp-drr}
\end{algorithm}

The algorithm realising our approach to combine RL-SP and DRR, is presented in
Algorithm~\ref{algo:rl-sp-drr}. Its data structures consist of an array of
priority queues, a list of active low priority queues needed for DRR and the
predefined high priority queue, which is the first queue with index 0 of all
priority queues. 

In the \texttt{ENQUEUE} procedure, the departure time of a packet from the
high priority queue is calculated based on the previous packet's departure
time and transmission time, which is expressed by the relation between packet
size and queue rate (line 7). To enqueue low priority packets, the departure
time is not needed, but the list of active queues must be maintained in order
to eliminate lookups of inactive queues, i.e. the ones that do not contain any
packets (line 11).

The \texttt{DEQUEUE} procedure starts with a check whether the departure time has come
for the packet of high priority queue. If it has, then the packet is sent from
this queue (line 14). Otherwise, the DRR algorithm is performed on low
priority active queues as described in~\cite{shreedhar1996efficient},
and the selected low priority queue by DRR is served (line 16). As soon as a
low priority packet is dequeued, the dequeue process again starts a check up
on the high priority queue.

\subsection{Implementation}

As proof-of-concept, the traffic manager has been implemented on a reference
P4 software switch---\bmv{} (Release 1.11.0), implemented in C++. The switch
supports input and output first-in-first-out (FIFO) buffers shared among all
ports and multiple intermediate (egress) buffers for traffic management.
Packet scheduling can be applied on the egress buffers. Buffering is performed
on the switch using threads synchronized with mutexes. As such, the
\texttt{receive} thread enqueues a packet into the input buffer; the
\texttt{ingress} thread then dequeues the packet and moves it to one of the
egress buffers. The egress buffers perform packet scheduling when packets are
dequeued by the \texttt{egress} threads from these buffers to the output
buffer. Eventually, the packets are sent to the output port from the output
buffer.

\bmv{} supports rate limited strict priority scheduling as an experimental
feature, having rate units as packets per second (pps). It is implemented by
defining an egress buffer as a C++ class, containing an array of
\textit{priority queues} with \textit{rate} property. In C++, a priority queue
stores the objects sorted according to their associated priorities, hence by
storing packets in the queue and assigning the departure time of packets as
such priority, the sorting of the packets from the earliest to the latest is
achieved. The departure time is calculated based on the \textit{rate} property
of the queue. When there are packets in the egress buffer, the \texttt{egress}
thread requests a packet from the buffer, which in turn polls the head packet
of each priority queue from most prioritised queue to the least. For each
queue, it compares the departure time of the head packet with the current
(wall) time and if the departure time has come, it dequeues the packet from
the queue and sends it to the \texttt{egress} thread, which further moves it
to the output buffer. The next lookup starts again from the most prioritized
queue.

We implemented several extensions of \bmv{} to realise the RL-SP-DRR traffic
manager. At first, the switch was extended to support rate limiting based on
bytes per second (Bps) units of measurement. This is needed, as the rate
limiting with \textit{pps} units assumes the packet size to be the size of
maximum transmission unit (MTU) leading to inaccurate data rates for packets
of smaller sizes. When the \texttt{ingress} thread enqueues the packet to the
egress buffers, the size of the packet is extracted and used to calculate the
packet's departure time for high priority queue based on the queue rate as
explained in the \texttt{ENQUEUE} procedure of Algorithm~\ref{algo:rl-sp-drr}.
For the low priority queues the rate limitation is implicit and not based on
the timing, but on quantum values, thus a wall time is assigned such that the
order of arriving packets will be preserved.

Secondly, we extended the strict priority scheduling code of \bmv{} in order
to follow the \texttt{DEQUEUE} procedure of Algorithm~\ref{algo:rl-sp-drr}. We
introduced a \textit{quantum} and a \textit{deficit counter} property to the
egress buffer and implemented the DRR~\cite{shreedhar1996efficient} algorithm
for low priority queues in the dequeue method of the egress buffer, while
keeping the polling order from the most prioritized queue to the least, such
that the priority queue will be served first, if its departure time has come.
Lastly, in order to configure parameters of the algorithm, such as priority,
rate and quantum values, respective metadata were introduced to the switch
model for programmability via P4 language and the parsing of these metadata
was added to the \bmv{}.

\section{Evaluation}
\label{sec:eval}

\subsection{Methodology}

For the evaluation, we use two evaluation scenarios (PROACTIVE and MPI), and for each of these, we compare the behaviour of our RL-SP-DRR scheduling algorithm against strict priority queueing (STRICT), and best-effort (NO). The two scenarios are defined below. For measuring network metrics and generating network traffic, we use the \texttt{iperf} tool. In our evaluation, UDP traffic was generated by the emulated hosts, such that the links got congested. 

\subsubsection{PROACTIVE scenario} In this scenario, all links are initially 
fully utilised by low priority hosts with configured packet scheduling. After
$N$ seconds, high priority traffic is injected for another $N$ seconds. In
this scenario, we use the UDP bandwidth distribution over time as a Key
Performance Indicator (KPI) for the network quality. The period of time is set
to 90 seconds ($N=90$) to make the bandwidth distribution clearer and see the
behaviour of the scheduling algorithms in the presence and absence of high
priority traffic.

\subsubsection{MPI scenario} In this scenario, low priority hosts are
generating traffic and high priority hosts are running an MPI application
that measures ping-pong latency by sending messages with different sizes and
100 times per size. The experiment is finished as soon as the MPI application
terminates. We use the MPI ping-pong latency between high priority hosts as a
KPI for the network quality.

\subsection{Set-up}

To deploy the evaluation setup, we use the mininet network emulator running
on a physical host with 4 cores of Intel i5-4590 and 16 GB of memory. To
support P4, the mininet switch ports need to be associated to the \bmv{} ports.

\begin{figure}
	\centering
	\includegraphics[width=\columnwidth]{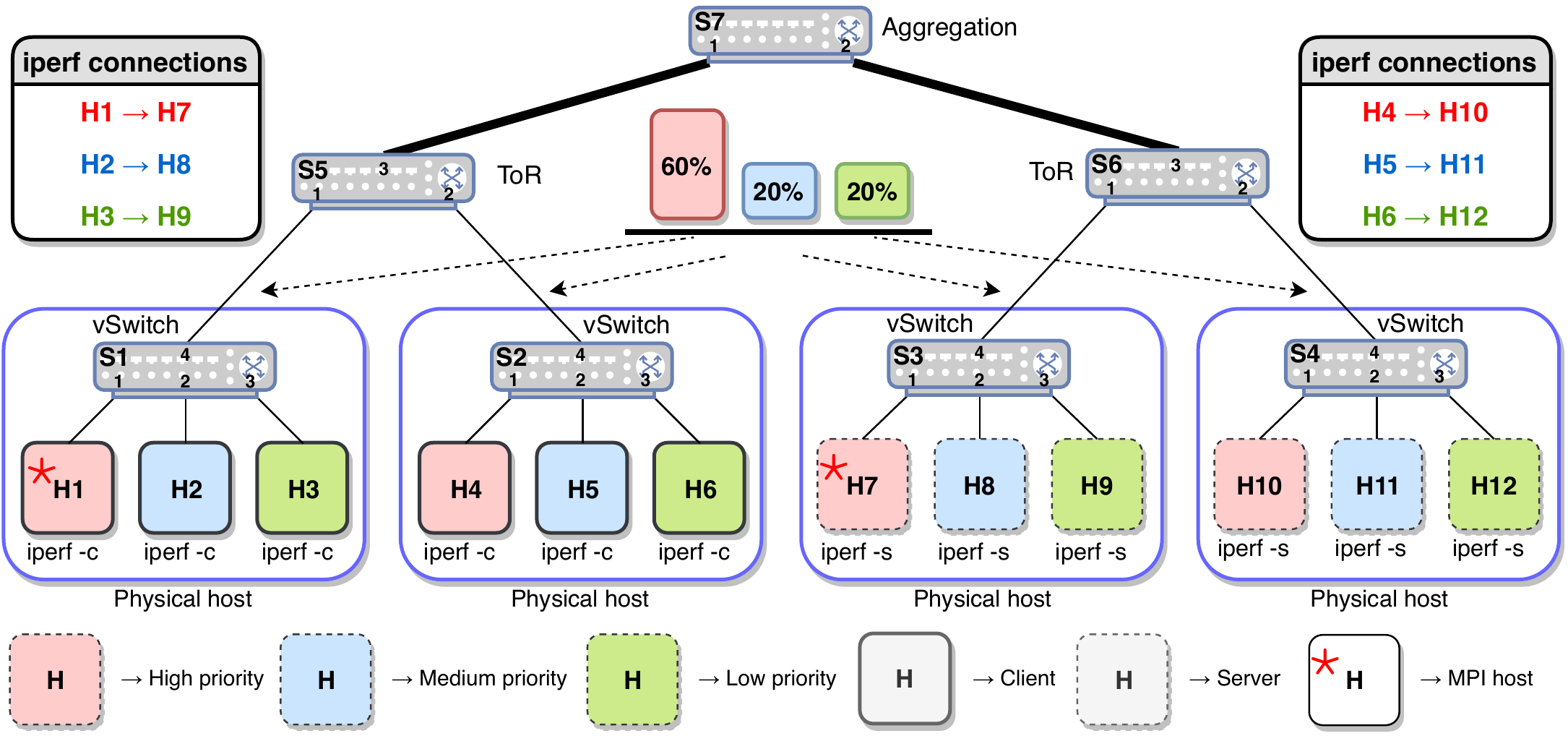}
	\caption{Tree (depth=2) with 12 hosts. Hosts H1-H6 are iperf-clients and hosts H7-H12 are iperf-servers. Links between physical hosts and ToR switch are congested and packet scheduling is applied to them for prioritization. The capacity of the links between ToR and aggregation is doubled}
	\label{fig:eval-setup-topo-3}
\end{figure}

The emulated network topology used for evaluation is a tree topology with the depth of
2, which simulates communication of leaf processes across an aggregation switch as
depicted in Figure~\ref{fig:eval-setup-topo-3}. In the figure, the branches
from the aggregation switch are separated into \texttt{iperf} clients (left branch) and
\texttt{iperf} servers (right branch). For the MPI test case, H1 and H7 are running the
MPI ping-pong latency measuring application. The set-up ensures that the links between 
physical hosts and ToR switches and between ToR switches and the aggregation switch are
congested. Since the latter source of congestion is related to the
over subscription problem at the aggregation tier, the capacity of these links
was doubled to sustain the load of two connected ToR switches. The
prioritisation scheme was then applied by the virtual switches, which is, 60\%
of available bandwidth is allocated to high priority. Further, we set the 
link capacities to 10 Mbps due to CPU constraints in the simulated software environments.

\subsection{Results}

The following sections summarise the results of the two evaluation cases.

\subsubsection{PROACTIVE scenario}

\begin{figure*}
	\centering
	\begin{minipage}[l]{0.65\columnwidth}
		\centering
		\includegraphics[width=\columnwidth,page=1]{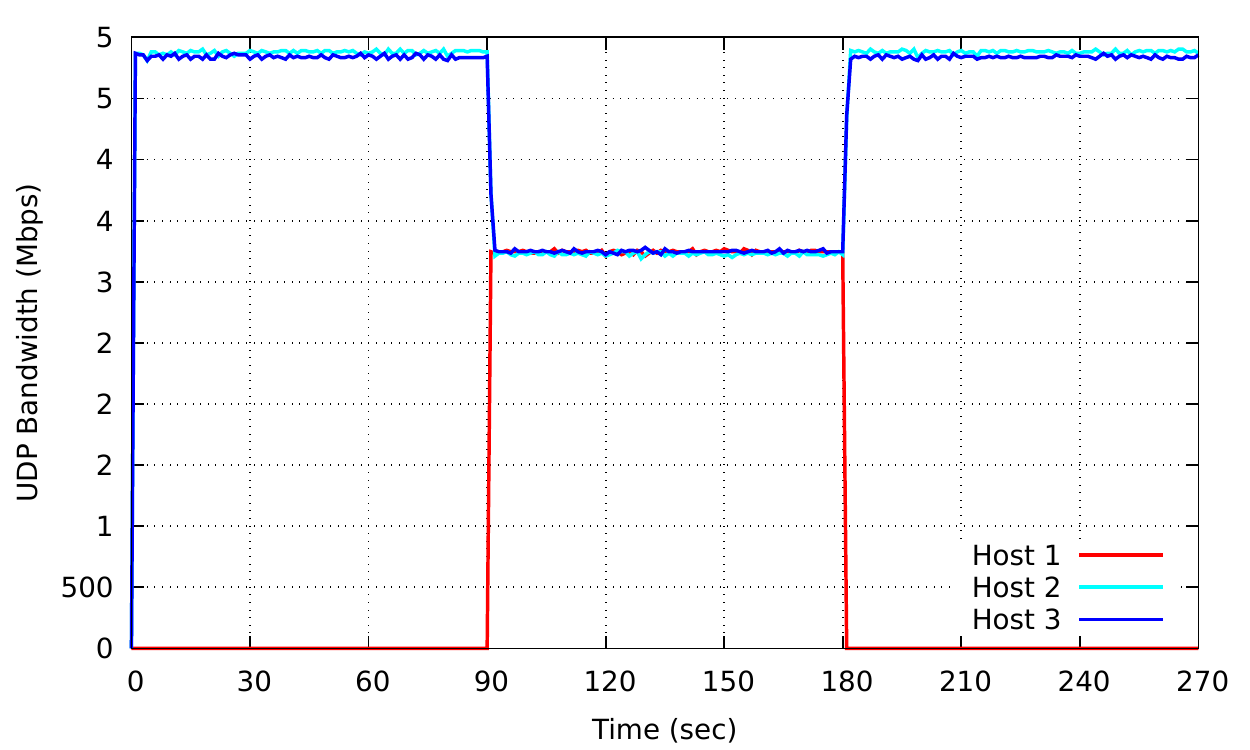}
		\caption{Bandwidth distribution. Best Effort}
		\label{fig:eval-t3-pro-def-udp-bw}
	\end{minipage}
	\hfill{}
	\begin{minipage}[r]{0.65\columnwidth}
		\centering
		\includegraphics[width=\columnwidth,page=1]{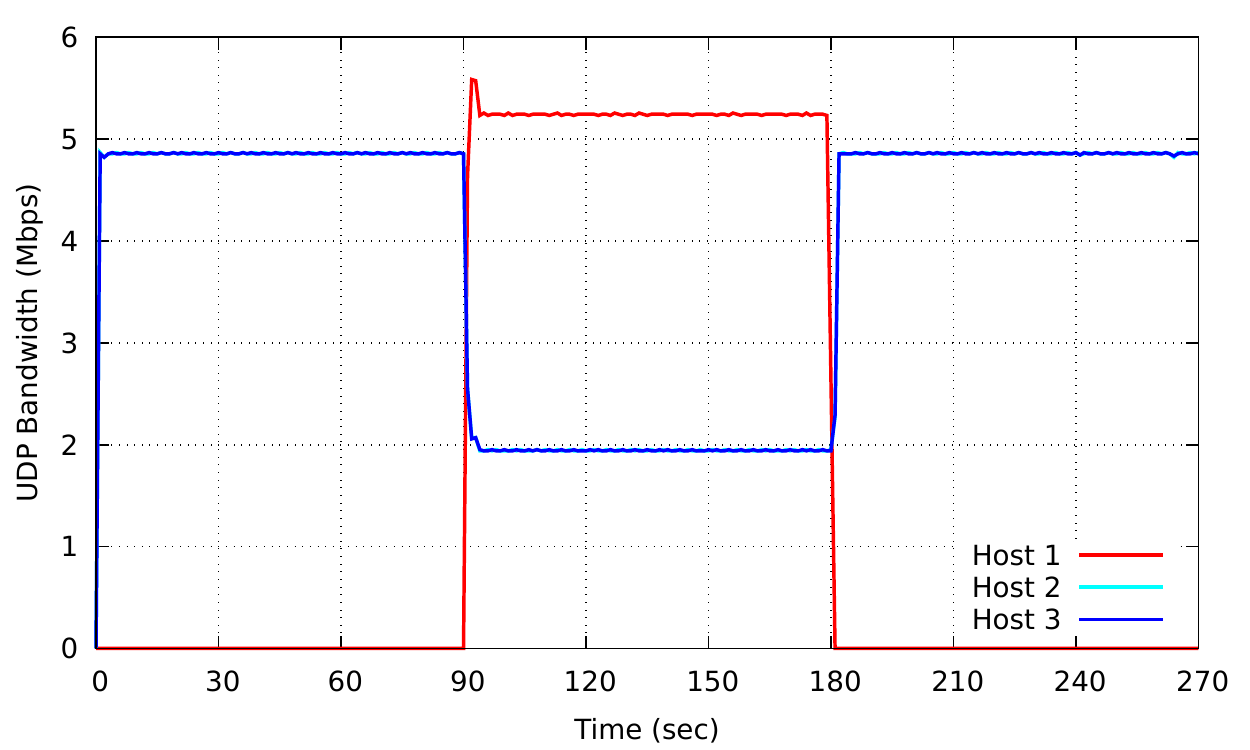}
		\caption{Bandwidth distribution. STRICT}
		\label{fig:eval-t3-pro-strict-udp-bw}
	\end{minipage}
	\hfill{}
	\begin{minipage}[r]{0.65\columnwidth}
		\centering
		\includegraphics[width=\columnwidth,page=1]{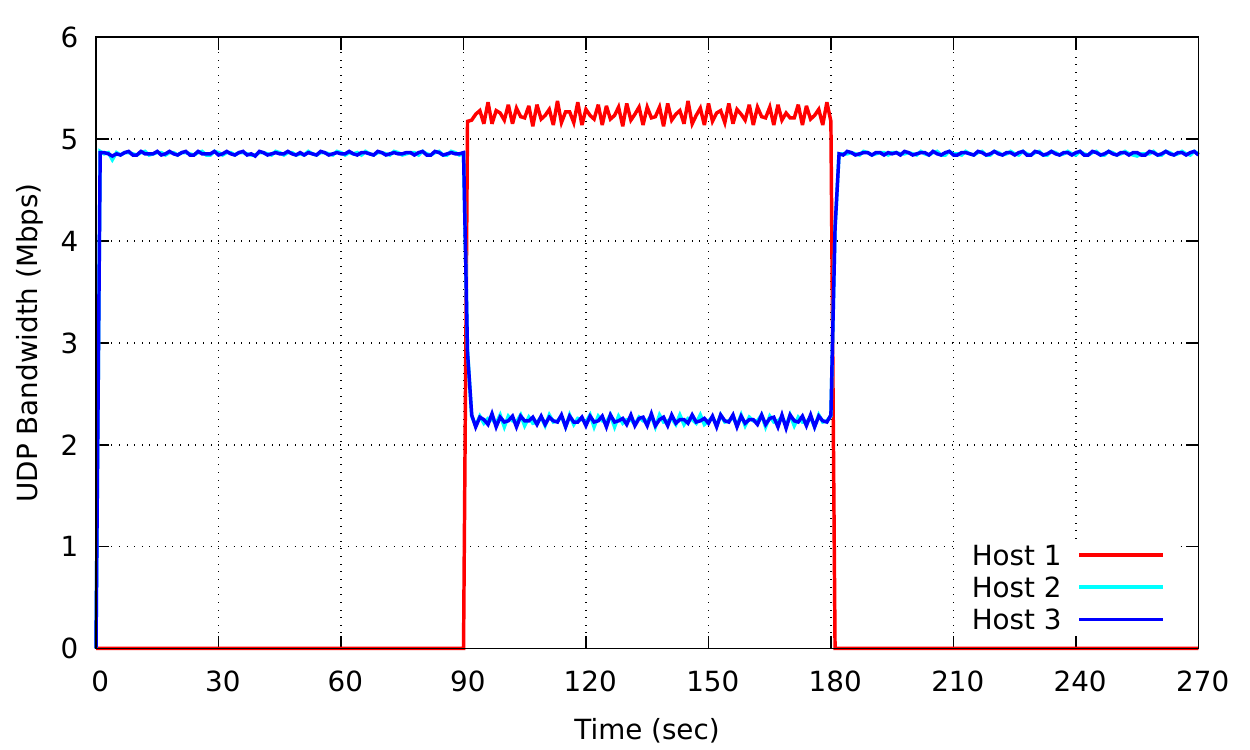}
		\caption{Bandwidth distribution. RL-SP-DRR}
		\label{fig:eval-t3-pro-drr-udp-bw}
	\end{minipage}
\end{figure*}

In this test case, NO demonstrates best-effort manner, as shown in figure
\ref{fig:eval-t3-pro-def-udp-bw}, where no traffic management is applied for
the uplink of the virtual switch. At the time of appearing, the high priority
traffic does not obtain higher bandwidth. In contrast, with enabled traffic 
management both STRICT and RL-SP-DRR prioritise the high priority traffic by providing
higher bandwidth share ($5.3$ Mbps). With UDP traffic (Figures~\ref{fig:eval-t3-pro-strict-udp-bw} 
and~\ref{fig:eval-t3-pro-drr-udp-bw}),
their behaviour is similar, but RL-SP-DRR slightly better utilises the link with
lower priorities.

\subsubsection{MPI scenarios}
\label{sec:eval-res-mpi}

\begin{figure}
	\centering
	\includegraphics[width=\columnwidth,page=1]{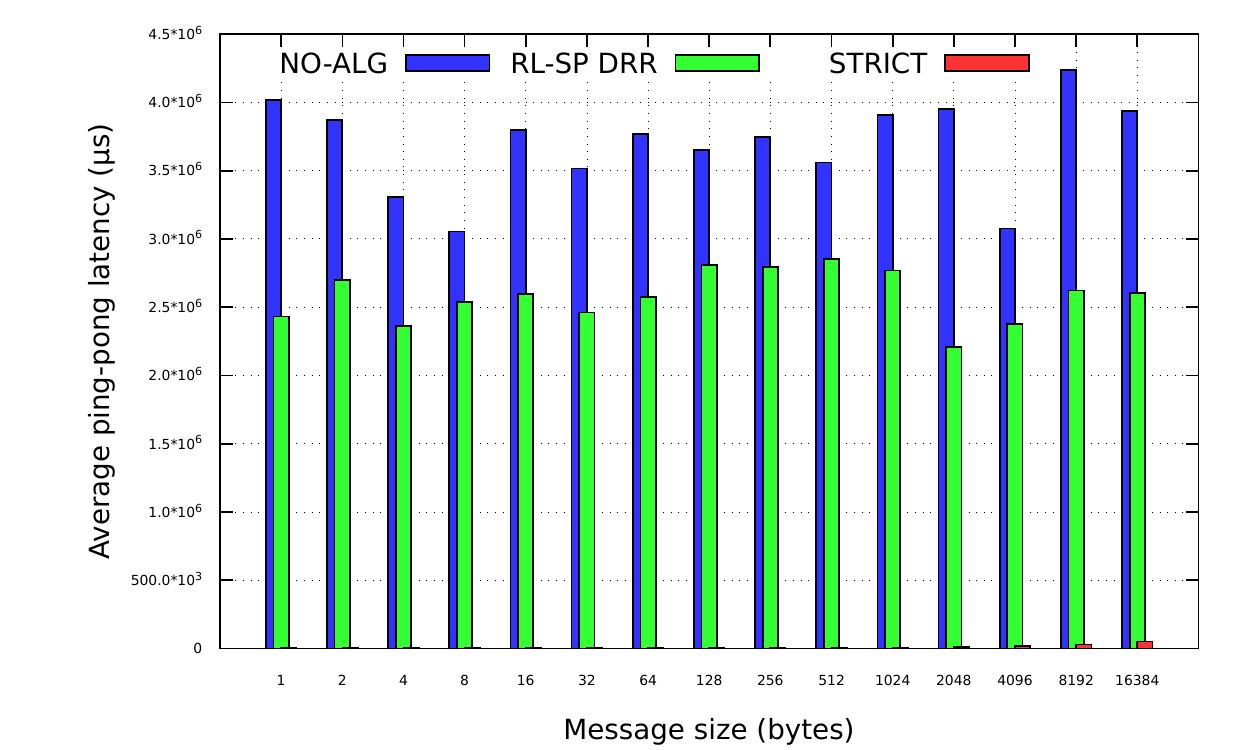}
	\caption{MPI scenario. Comparison of ping-pong latencies}
	\label{fig:eval-t3-mpi-udp}
\end{figure}

In the MPI test case, the prioritisation improves the performance of MPI
communication, as shown in Figure~\ref{fig:eval-t3-mpi-udp}, when compared to
the case without traffic management. For over all message sizes, the latencies of RL-SP-DRR and STRICT are lower, e.g. for message size of 16384 bytes, the latencies are $2.6 s$ and $51 ms$ respectively on average, while NO has $3.9 s$. 

\begin{figure}
	\centering
	\includegraphics[width=\columnwidth]{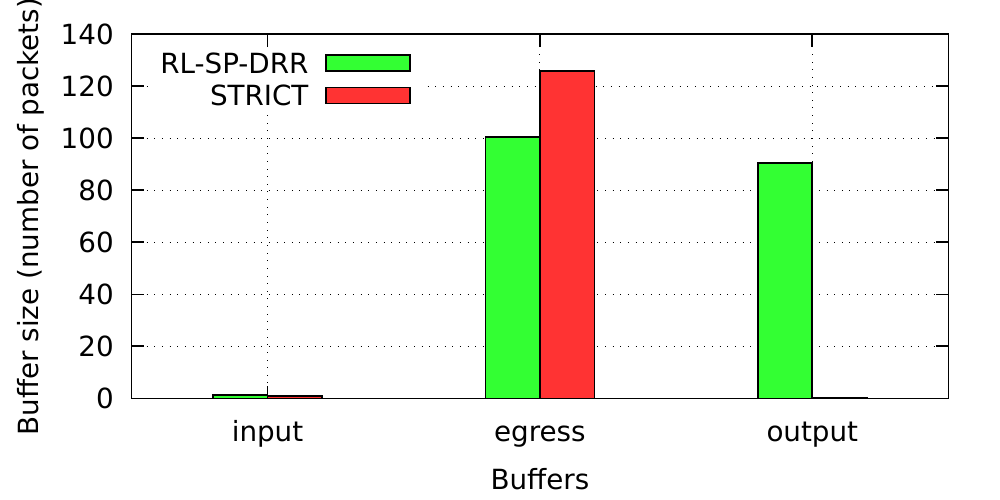}
	\caption{Input, egress and output buffers occupancy for strict priority and RL-SP-DRR}
	\label{fig:impl-buff-space}
\end{figure}

There is a latency improvement for RL-SP-DRR, however, it is outperformed by
STRICT. In order to justify these results, buffer occupancy of the switch
was inspected and is presented in Figure~\ref{fig:impl-buff-space}. STRICT
does not allow the output buffer to be full due to explicit rate limitation imposed on low priority egress buffers,
thus giving virtually no delays for high priority packets. RL-SP-DRR, on the
other hand, overruns the output buffer with low priority packets due to implicit
rate distribution based on quantum values. Hence, the queue delays for RL-SP-DRR is
higher. Therefore, random early detection (RED) or some of its flavour should be
applied on the output buffer, such that the buffer space will be free for high
priority traffic.

\subsection{Discussion}

When comparing traffic management accomplished with STRICT and RL-SP-DRR, we see that the 
shaping of UDP traffic is comparable in terms of bandwidth allocation. However, it is clear that the prioritisation done with STRICT
outperforms RL-SP-DRR in terms of latency due to the rate limitation and explicit refraining of low priority
egress buffers, which keep the output buffer from overrun. The major issue, however, with STRICT is its
scalability. To elaborate this, we calculated the number of control plane commands for
bandwidth allocations:

In the evaluation we imposed the condition that all links are fully utilised. Assume that there are $N$ virtual switches and $M$ leaf processes per switch. Assuming that (for simplicity), only one such process has high priority ($P = 1$). The initial network pre-configuration requires that the control plane needs to send $N*M$ commands to shape the traffic for all processes including the prioritised one. For RL-SP-DRR, in contrast, the traffic is implicitly shaped using predefined quantum values, thus only the commands for prioritisation are needed: $N*P=N$, for $P=1$. So there is $M$ times less commands to be sent by the control plane. Assuming 30 processes per physical host, 30 times less commands can be achieved.

The proactive bandwidth allocation is not possible with STRICT, if the
shaping is required for low priority traffic and the links should be utilised
entirely (cf. PROACTIVE case). Therefore, if the prioritisation scheme is proactively enabled, then adjustments of the rates are needed, but only for low priorities. Here, the number of commands is decreasing to $N*(M-P)=N*(M-1)$, for $P=1$. However, with RL-SP-DRR, traffic of any priority can be shaped and prioritised proactively utilising the links without sending any further control plane commands. 

Hence, if traffic shaping is not required for lower priorities, the best-effort approach is suitable, and the number of processes as well as their priorities are static, then STRICT can be used for bandwidth guarantee. In all other scenarios, RL-SP-DRR is superior. Despite that, it can be further improved with better buffer management.

\section{Conclusions and Future Work}

In this work, we presented a traffic manager with the goals of complete link utilisation, prioritisation and scalability using rate limited strict priority with deficit round-robin (RL-SP-DRR) packet scheduling algorithm. We implemented this algorithm in the \bmv{} P4 software switch relying on the flexible metadata-based interface for its configuration. In this way, parameters of the algorithm, e.g. priority, rate and quantum, can be provisioned by a pipeline table, which, in turn, is populated by the control plane. Hence, programmable traffic management using a standard control plane API is achieved.

Traffic management with RL-SP-DRR benefits in scalability and
link utilisation. Much fewer control plane commands are required when
prioritisation needs to be made at runtime. Moreover, no control plane
interactions are needed to sustain link utilisation in contrast to dynamic
rate distribution with strict priority, once a preliminary prioritisation scheme
is established. Evaluation on the \bmv{} switch shows that the approach is 
superior to existing approaches in high utilisation environments. As future work, we plan to improve the buffer management of the switch to further 
reduce queue delays. Finally, the performance shall be tested on hardware solutions since current results demonstrate the concept only on software-based emulation.

\section*{Acknowledgements}

The research leading to this paper has received funding from the European Union's Horizon 2020 research and innovation programme under grant agreement No 732667 (RECAP) and 732258 (CloudPerfect), as well as from the government of the federal state of Baden-W{\"u}rttemberg, Germany for the bwCloud and bwNet100G+ projects.

\bibliographystyle{IEEEtran}
\bibliography{IEEEabrv,references}

\end{document}